# High Power, High Energy Cyclotrons for Decay-At-Rest Neutrino Sources: The DAEδALUS Project

Jose R. Alonso for the DAEδALUS Collaboration
*LNS-MIT, Cambridge MA 02138, USA*

Neutrino physics is a forefront topic of today's research. Large detectors installed underground study neutrino properties using neutrino beams from muons decaying in flight. DAEδALUS looks at neutrinos from stopped muons, "decay at rest" (DAR) neutrinos. The DAR neutrino spectrum has effectively no electron antineutrinos (essentially all π- are absorbed), so a detector with free protons is sensitive to appearance of nu-e-bar oscillating from nu-mu-bar via inverse-beta-decay (IBD). Oscillations are studied using sources relatively near the detector, but which explore the same physics as the high-energy neutrino beams from typical Long Baseline experiments. As the DAR spectrum is fixed, the baseline is varied: plans call for 3 accelerator-based neutrino sources at 1.5, 8 and 20 km with staggered beam-on times. Compact, cost-effective superconducting ring cyclotrons accelerating molecular hydrogen ions ($H_2^+$) to 800 MeV/n with stripping extraction are being designed by L. Calabretta and his group. This revolutionary design could find application in many ADS-related fields.

## 1. Introduction

The sensitivity of neutrino experiments can be characterized by the product of the neutrino flux at the site of the detector and the volume of the detector. Consequently, detectors are built as large as can be afforded and are located at relevant locations with respect to neutrino sources. The geometry and sensitivity generally remain fixed throughout the lifetime of the experiment. However, if cost-effective sources of neutrinos – with adequate flux – were available, new generations of experiments could be planned by placement of such sources close to existing detectors. Such an opportunity would allow for new uses for very large detectors designed for long-baseline studies. The DAEδALUS experiment is based on this premise, with planned placement of three compact decay-at-rest neutrino sources at calculated distances from the large water-Cherenkov detector planned for the Sanford Laboratory at the Homestake Mine in the Black Hills of South Dakota [2]. This paper will describe the experiment, and its significant complementarity with measurements using the LBNE neutrino beam from Fermilab [3], then address the accelerator requirements to produce the required neutrino fluxes and possible accelerator technologies that would meet these requirements. In particular, a promising new concept for compact, high-current cyclotrons accelerating molecular hydrogen ($H_2^+$) beams [1] will be described.

## 2. Neutrino Oscillation Experiments

Measurement of oscillation parameters for neutrinos can shed light on CP violation in the neutrino sector, and, as has been recently shown, possibilities of new physics in the difference between behavior of neutrinos and antineutrinos in short-baseline experiments. Starting with a beam of muon neutrinos or antineutrinos, the probability of oscillating into an electron neutrino or antineutrino is given in Equation (1).

$$P(\nu_\mu \to \nu_e) = \sin^2\theta_{23} \sin^2\theta_{13} \sin^2\Delta_{31} \qquad (1)$$
$$\pm \sin\delta \; \sin2\theta_{13} \sin2\theta_{23} \sin2\theta_{12} \; \sin^2\Delta_{31} \sin\Delta_{21}$$
$$+ \cos\delta \; \sin2\theta_{13} \sin2\theta_{23} \sin2\theta_{12} \; \sin\Delta_{31} \cos\Delta_{31} \sin\Delta_{21}$$
$$+ \cos^2\theta_{23} \sin^2\theta_{12} \; \sin^2\Delta_{21}$$

where the ($\sin \delta$) sign is positive for $\overline{\nu_\mu} \to \overline{\nu_e}$, and negative for $\nu_\mu \to \nu_e$. $\delta$ is the CP violation phase, $\theta_{ij}$ are the neutrino mixing angles, and

$$\Delta_{ij} = \Delta m^2_{ij} L/4E_\nu \qquad (2)$$

shows the dependence on mass splitting, baseline length L and neutrino energy $E_\nu$.

## 2.1. Long Baseline Measurements

Conventional Long Baseline experiments direct a high-energy beam[1] of protons at a target, producing a spray of secondary particles: pions and kaons mostly. A focusing horn or lens immediately following the target provides a means of concentrating one or the other sign of secondary particles along the beam axis, and a subsequent long (~1 km) large-diameter tube, evacuated if possible, provides for decay in flight of the pion or kaon leading to kinematically focused neutrinos. Remaining particles exiting the back of the decay pipe stop in the rock, and do not contribute to the neutrino beam. By orienting the primary proton beam towards a distant detector (> 250 km away), events seen in the detector will most likely have originated from neutrinos produced from these protons. The focusing horn can select positive or negative particles (neutrinos or antineutrinos respectively) based on the direction of the current flow. Switching between neutrino and antineutrino modes is accomplished by simply reversing this current.

As the sign of the (sin δ) term in Equation (1) depends on whether the source is neutrinos or antineutrinos, experimental investigation of CP violation using the conventional Long Baseline technique requires running in both modes to observe the difference. Experiments to date have measured $\nu_\mu$ and $\overline{\nu}_\mu$ disappearance, difficult measurements because of normalization, efficiency and calibration uncertainties. Of particular note is the higher background and lower sensitivity in the antineutrino mode. Lower π- yield, higher π+ contamination and matter effects from traversals of long distances through rocks all contribute to poorer statistics in the antineutrino mode. Nonetheless, T2K [4] and MINOS [5] have yielded good results, and new long baseline experiments are beginning data taking (ICARUS, OPERA) or are in planning (LBNE).

Specifically, LBNE, the Long Baseline Neutrino Experiment [3] is being planned based on a new beamline at Fermilab directed to the Sanford Lab in Lead, South Dakota, 1300 km away. Originally conceived as a 300 kiloton water-Cherenkov detector, the actual technology and size are still being evaluated. Liquid argon (~17 kT) or water Cherenkov (200-300 kT), or even possibly liquid scintillator (35 kT) would all provide the sensitivity desired for the LBNE experiment.

## 2.2. DAEδALUS Configuration

Equation (1) provides another avenue to measuring CP violation, through the $\Delta_{ij}$ terms. As expressed in Equation (2), $\Delta_{ij}$ depends on the mass splitting, neutrino energy and baseline length. DAEδALUS plans on making measurements at different baseline lengths, while keeping all other parameters the same.

"Decay At rest Experiment for $\delta_{cp}$ At the Laboratory for Underground Science" (DAEδALUS) aims to produce π+ with low enough energy that these particles will stop in the target prior to decaying. At approximately 800 MeV, not far away from the production threshold energy of just below 600 MeV, π- production is substantially lower, and most all π- produced are absorbed in nuclei prior to decaying. Calculations for a graphite target indicate the suppression of π- is approximately $10^4$. As seen in Figure 1, the spectrum of neutrinos from the decay-at-rest (DAR) of π+ (and subsequent μ+ decay) is devoid of electron antineutrinos, and has an average energy of the muon antineutrinos of about 40 MeV.

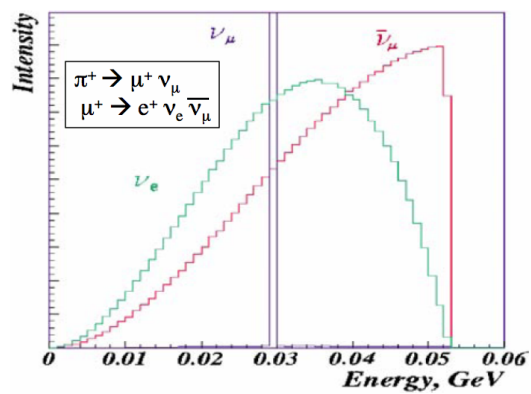

Figure 1: Neutrino spectrum from π+ decaying at rest. The spectrum is devoid of electron antineutrinos.

Referring to Equation (1), neutrino parameters for the DAR experiment are fixed: one cannot change between neutrino and antineutrino, and the energy is fixed by nature. However, one has the ability to generate neutrinos at different baseline lengths, and thus obtain data relevant to $\delta_{cp}$. In fact, the DAEδALUS experiment proposes three neutrino sources, as shown in Figure 2, with baselines of 1.5 km, 8 km and 20 km. The 1.5 km station provides flux normalization and calibration data, the 8 km site is at the π/4 point of the oscillation spectrum and the 20 km site at the maximum, π/2, point (see Figure 4).

---

[1] Beams at Fermilab are 8 and 120 GeV, at JParc 30 GeV and at CERN 400 GeV.

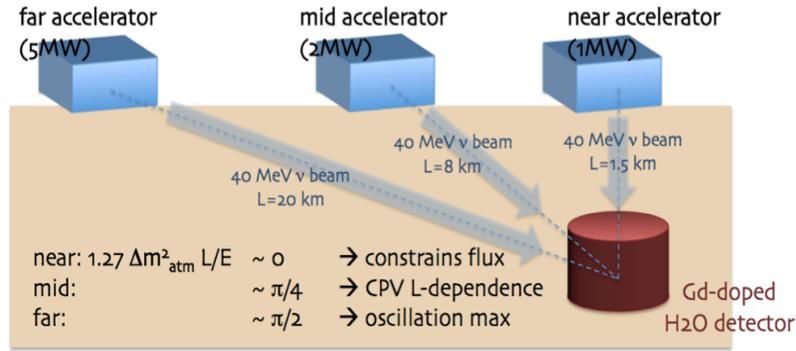

Figure 2: The DAEδALUS configuration with 3 DAR neutrino sources.

### 2.3. Detecting Neutrinos

A detector with a high fraction of free protons, such as a water-Cherenkov or liquid-scintillator detector, is particularly well-suited to detect electron antineutrinos, via the inverse-beta-decay process, as illustrated in Figure 3. The electron antineutrino strikes a proton, produces a relativistic positron with a cone of Cherenkov radiation, and a free neutron. The neutron will be eventually captured, a reaction enhanced if the detector is doped with a substance such as Gd with a high capture cross section and for which the {n,γ} capture reaction releases a substantial amount of energy. This capture occurs typically a few microseconds later, yielding a very clean and unique delayed-coincidence signature for the event.

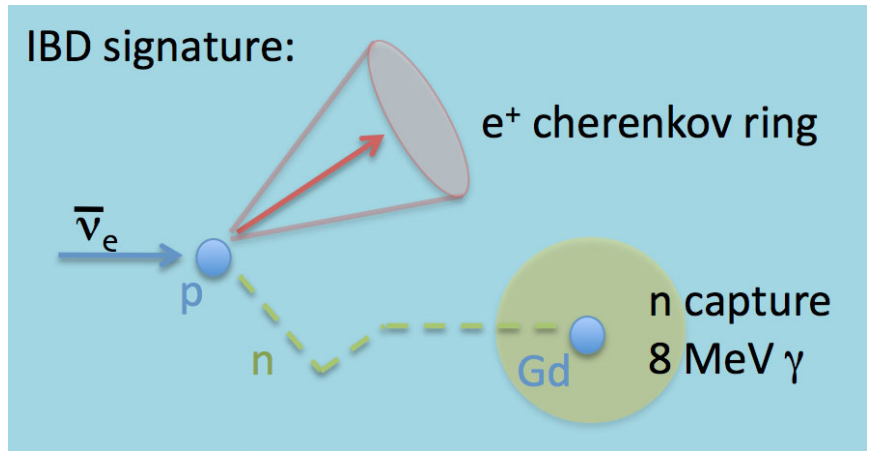

Figure 3: Inverse Beta Decay (IBD) process for absorption of an electron antineutrino in a detector with large amounts of free protons. The prompt Cherenkov cone is followed a few microseconds later by the absorption of the free neutron, producing a clean delayed-coincidence signature for the event, easily distinguishable from background.

### 2.4. The DAEδALUS Experiment

As the DAR spectrum contains no electron antineutrinos, the detection of these antineutrinos is a measure of oscillation from muon antineutrinos. This "appearance" measurement provides greater sensitivity and more background rejection than the conventional long-baseline disappearance measurement.

Though the neutrino energies and baselines are substantially different from the LBNE configuration, both explore the same regime of oscillation physics due to similarity of the L/E ratios. LBNE baseline is $10^3$ km, and neutrino energy is ~2 GeV, while the DAEδALUS baseline is 20 km and neutrino energy 40 MeV.

Figure 4 illustrates the basic oscillation concept. Muon (anti)neutrinos start off at the left end; measurements made along the flight path have a probability of observing an electron (anti)neutrino. The amplitude of the oscillation probability is related to the mixing angle $\theta_{ij}$, the wavelength to $E / \Delta m_{ij}^2$. The 8 and 20 km points for the DAE$\delta$ALUS measurements are indicated. From Equation (1) $\delta$, the CP-violating phase, also enters into the amplitude of the oscillation probability.

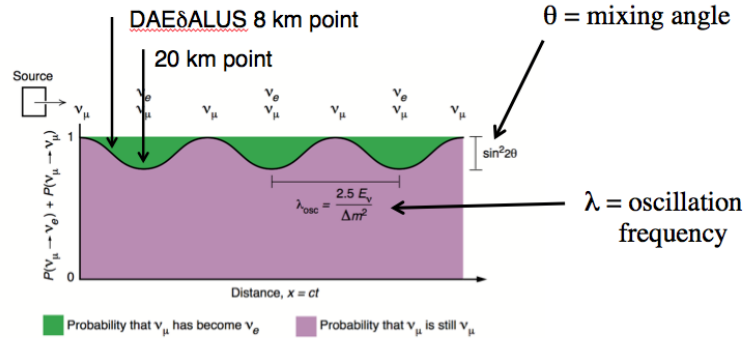

Figure 4: Neutrino oscillation probability. Amplitude depends on mixing angle as $\sin^2 2\theta$ (as well as on $\delta$, the CP-violation phase); wavelength is dependent on the ratio of the neutrino energy to the mass-squared difference.

Two parameters in Equation (1) are currently not well known: $\theta_{13}$ and $\delta$. Current limit on $\sin^2 2\theta_{13}$ is less than .05, and two experiments currently taking data (Double CHOOZ [6] and Daya Bay [7]) are expected to provide better answers within about 5 years. In the interim, however, sensitivity estimates for experiments can be evaluated with these unknowns as free parameters, generating what are known as "jelly-beam" plots described in the following section.

## 2.5. Sensitivity Comparisons

Figure 5a shows a schematic for a sensitivity plot demonstrating the methodology of assuming values for $\theta_{13}$ and $\delta$, and estimating the 1$\sigma$ and 2$\sigma$ contour plots for measurements made in a given experiment. In this formalism, the smaller the "jelly bean" the more precise the measurement. Figure 5b shows this plot for the DAE$\delta$ALUS experiment operating for a 10-year period, assuming a 300 kT water-Cherenkov detector. Figure 5c shows a plot for the LBNE experiment, operating also for a 10-year period, 5 years in neutrino mode and 5 years in antineutrino mode.

Figure 5d shows the results of combining the datasets for the two experiments. Considering that both use the same detector, and can run simultaneously, the symbiosis between the two measurements provides a very substantial improvement in the experimental sensitivity!

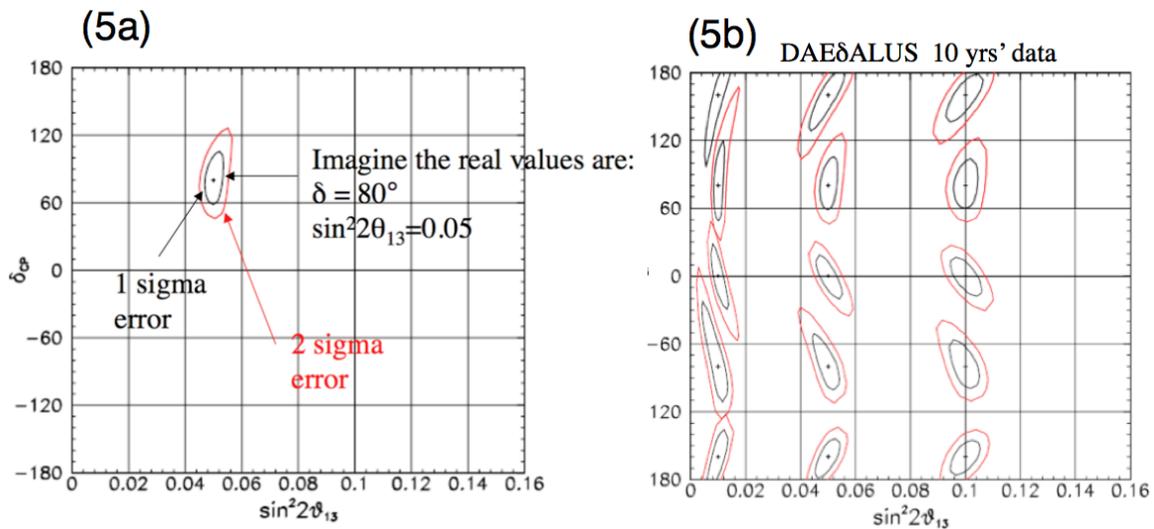

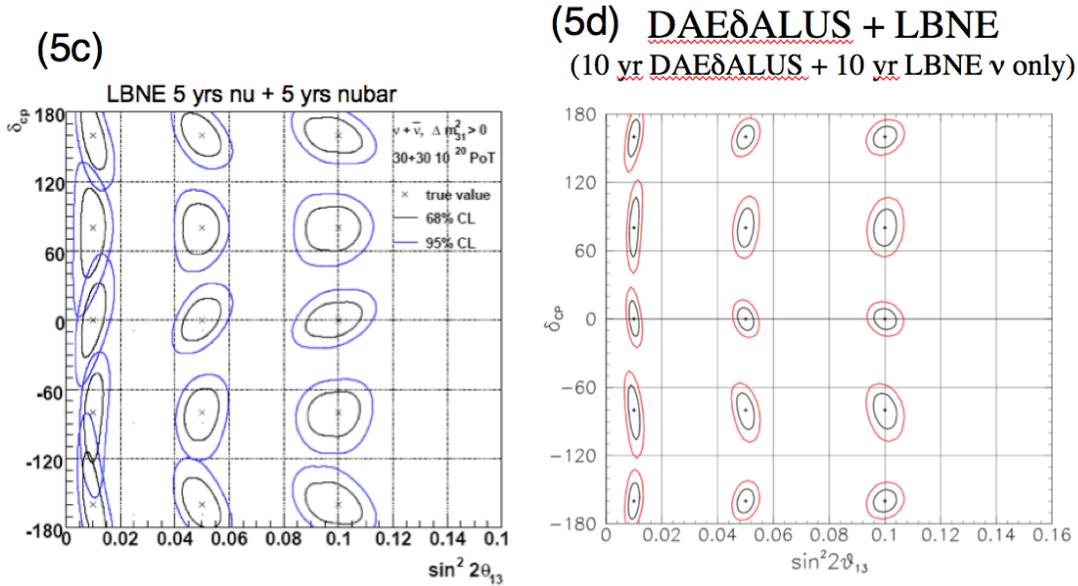

Figure 5: "Jelly Bean Plots" for DAEδALUS and LBNE experiments. a) Assume values for the two unknown parameters, $\theta_{13}$ and $\delta$, evaluate likely measurement result for each experiment, with contours for $1\sigma$ and $2\sigma$. The smaller the "jelly bean" the more precise the measurement. b) Assumed 10-year running for the DAEδALUS experiment alone. c) Assume 10-year running for LBNE, with 5 years each in neutrino and antineutrino modes. d) Combining the datasets from DAEδALUS and LBNE provides a very large improvement in sensitivity.

Both experiments can take data simultaneously, as the DAEδALUS sources run essentially continuously, while the LBNE pulses last a few microseconds and come every 3 seconds or so. This very low duty factor does not detract from the live-time for DAEδALUS.

The symbiosis is particularly relevant should the value of $\theta_{13}$ be very small. Figure 6 emphasizes this, with families of curves showing relative sensitivities for different configurations and different sources of neutrinos aimed to the large water-Cherenkov detector at the Sanford Lab.

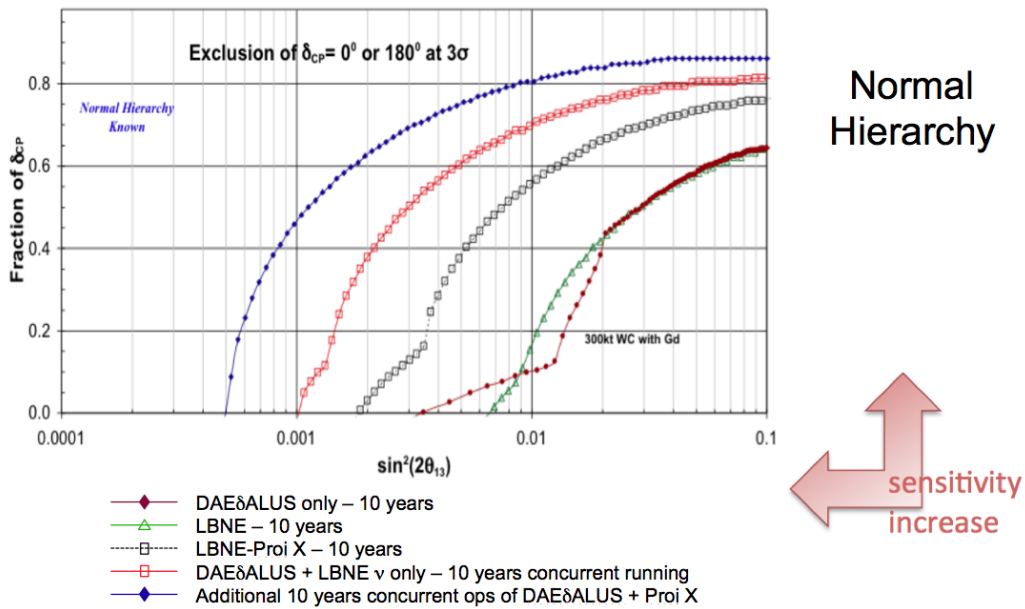

Figure 6: Comparison of data sensitivity for DAEδALUS and different running modes of LBNE. The curves bound the area excluded (to the left of the curves) for $\sin^2 2\theta_{13}$ and $\delta_{CP}$, where measurements can be differentiated from $\delta_{CP} = 0°$ or $180°$ at $3\sigma$. Plot is for normal mass hierarchy; the plot for inverted mass hierarchy is quite similar. Project X refers to a linac upgrade program contemplated for Fermilab that would increase the power on target by about a factor of 5 or more.

In these sensitivity comparisons, the total flux of neutrinos estimated for DAEδALUS operation is $4 \times 10^{22}$ per megawatt-year of 800 MeV protons, and thus sets the DAEδALUS accelerator requirement: to match the sensitivity of the planned basic LBNE configuration.

### 2.6. Sterile Neutrinos: New Experimental Possibilities

Recent very interesting results from antineutrino running at MiniBooNE, which appear to confirm earlier LSND results showing oscillation at short baselines, while neutrinos at the same configurations do not seem to show oscillations [8], are leading to speculation for existence of one or more sterile neutrinos which can add required flexibility to the mixing matrix necessary to fit the new diversity of data. Best fits to all existing data, including the recently-noted reactor antineutrino anomaly (resulting from recalibration of the flux of antineutrinos from the basic reactions in reactor fuel elements) [9], point to a mass difference, $\Delta m^2$ of the order of 1 eV$^2$. This compares to $\Delta m_{12}^2 \approx 8 \times 10^{-5}$ and $\Delta m_{23}^2 \approx 3 \times 10^{-3}$ eV$^2$.

With such a large mass difference, oscillation wavelengths for DAR spectra would be much shorter, the $\pi/2$ point being at about 40 meters. Figure 7 shows a possible Sterile Neutrino experiment using the same DAEδALUS concept, however here the scale of target placement is such that three targets could all be serviced sequentially by a single accelerator.

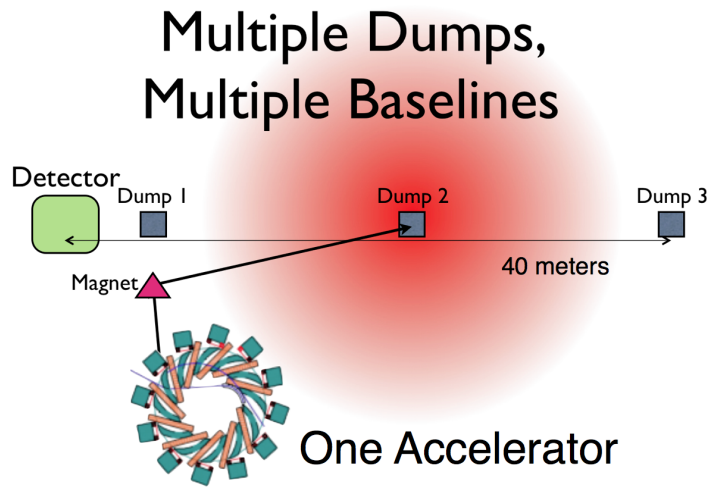

Figure 7: Configuration for a possible sterile-neutrino experiment, where the very short baseline (~40 meters for $\pi/2$ oscillation point) could enable a single accelerator servicing three different target stations.

### 3. Accelerator Requirements

With such compelling physics, the important remaining question is whether these neutrino sources can be built for an affordable cost. To assess this question one must understand the requirements for the accelerator, and how these might be met. Table I summarizes the basic parameters for the accelerator.

Table I: Accelerator requirements for DAEδALUS neutrino sources

| Beam on Target | Protons |
|---|---|
| Proton Energy | ~800 MeV |
| Duty Factor | 20% |
| Average Power | 1/2/5 MW |
| Peak Power | 5/10/25 MW |
| Neutrino flux | $4 \times 10^{22}$ ν/MW-protons / year |
| Acceptable beam loss | <200 W @>100 MeV |

Protons are by far the best particle available for producing pions. An examination of production cross sections, available flux and contaminants using any other projectile quickly confirm this.

Optimum proton energy is about 800 MeV, as can be seen in Figure 8. The pion production threshold is slightly below 600 MeV, the extra 200 MeV allows production to continue while the protons are losing energy in a thick target, an energy loss of 200 MeV corresponding roughly to one nuclear mean free path in the graphite target material most likely to be used.

Higher energy begins opening reaction channels above the delta resonance, and also increases the production of π- and the chances that these will decay in flight before being absorbed, yielding a contaminating source of electron antineutrinos.

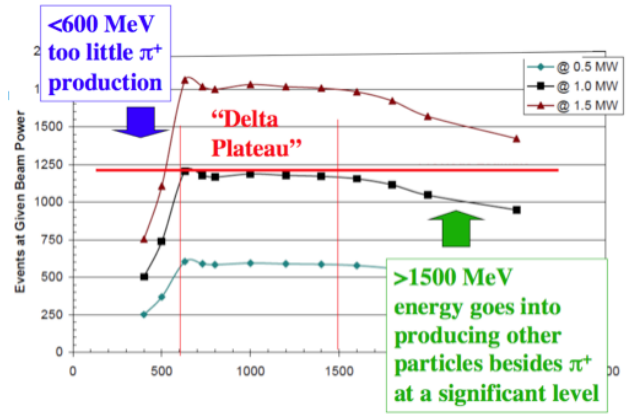

Figure 8: $\pi^+$ yield vs proton energy at constant beam power.

Duty factor considerations are important. The inverse beta decay process in the detector does not contain unambiguous directionality information, so an event observed in the detector cannot be traced back to a particular source. Consequently, only one source can be on at any given time so events can be tagged according to which source was active at the time the event was recorded. Figure 9 shows an example of an operating scheme for the three sources. The 20% duty factor for each, with 40% in which no sources are producing neutrinos is required for good background measurement.

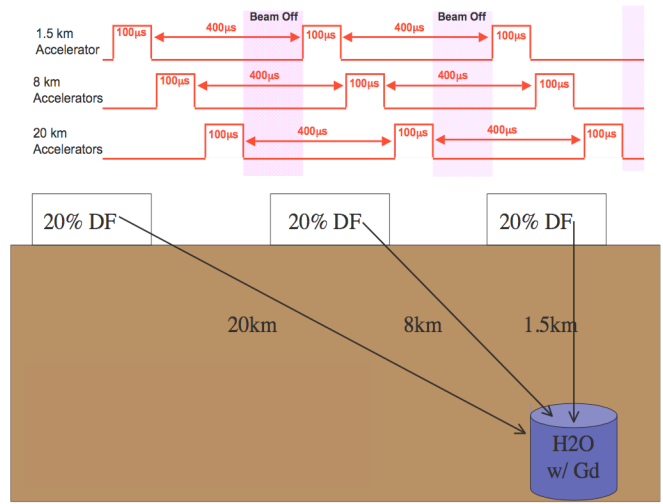

Figure 9: Example of beam-on/off cycle for each neutrino source, to enable tagging of events by time. The actual time for each beam-on period is arbitrary, should just be substantially longer than the muon lifetime of 2 μsec. This time period can be optimized based on accelerator and target engineering considerations.

The consequence of this is requiring higher instantaneous power output from each accelerator. The power levels for each of the stations of 1/2/5 MW, shown in Figure 2 reflects the flux necessary to accrue statistics from each source commensurate with the accrual of data from the LBNE experiment, these are the *average* power levels. As each source is on for only 1/5 of the time, the *instantaneous* powers will be a factor of 5 higher.

These power levels present a formidable challenge for accelerator designers, primarily in the requirement to control and minimize beam losses at high energies. The constraint relates to the survival of the accelerator components against thermal damage from stray beam, and that radiation levels from activation via beam loss are low enough to allow hands-on maintenance. Experience at PSI [10] indicate that beam losses at high energies (where activation is most severe) must be kept below 200 watts in each cyclotron vault. Considering that the total beam power of the PSI Ring Cyclotron is over 1 MW, the specification for uncontrolled beam loss is 2 parts in $10^4$. This specification is absolute, not a ratio of beam power, so if the average power from the accelerator is 5 MW the loss specification becomes up to 5 times tighter.

## 4. Technology Options

Can such a machine be built? As is seen in Figure 10 the challenges are daunting, but probably not impossible.

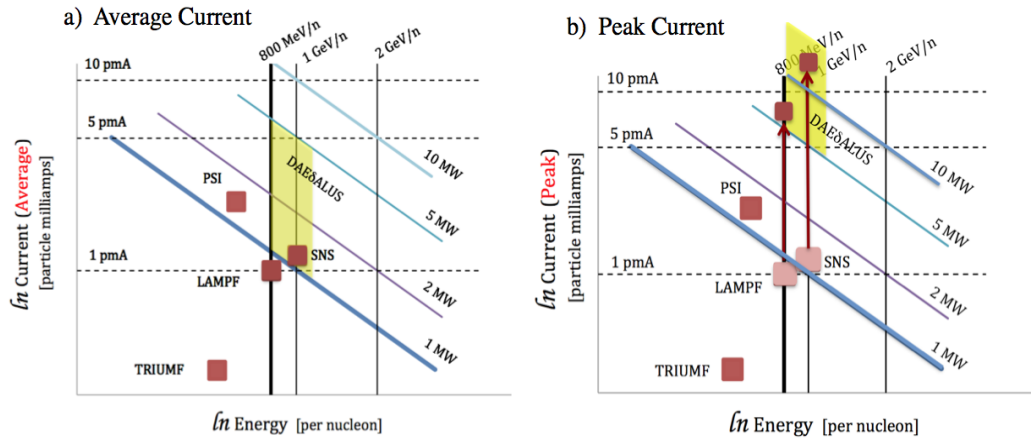

Figure 10: log(I) vs log(E) plots of existing accelerators and DAEδALUS requirements, diagonal lines are constant power. a) Average power and, b) peak power. SNS runs at 6% duty factor with average power of 1 MW, LAMPF at 12% with about 0.8 MW of average power. PSI is a cyclotron facility running at 100% duty factor.

Linear accelerators, in particular using superconducting technology, are certainly capable of meeting the DAEδALUS requirements. Efficiencies are high, apertures large and extracting beam at high energy requires no special gymnastics, so meeting the stringent beam-loss requirements is not difficult. However, these machines are very expensive and large; building the DAEδALUS experiment around 3 such linacs would be difficult to sell.

The question is whether there might be another option that would meet the specifications, and be more compact and substantially less expensive.

## 5. Cyclotron Alternatives

Cyclotron technology is very mature, from small isotope-producing units accelerating H- beams to large machines, both superconducting and normal-conducting for heavy-ion or high-current research applications. Particularly relevant to the present project are the Ring Cyclotron at the Paul Scherrer Institute in Villigen Switzerland [11] described in section 5.1, and the superconducting Ring Cyclotron (SRC) [12] at RIKEN, Wako, Japan, which although designed for high-energy highly-stripped heavy-ion beams, represents an engineering "proof-of-principle" design for a cyclotron magnet such as discussed in section 5.2, and possibly applicable for DAEδALUS.

### 5.1. The PSI Proton Ring Cyclotron

With an energy of 590 MeV and beam current of 2.2 mA, the PSI Ring Cyclotron is currently the world's most powerful accelerator in this energy range, delivering 1.3 MW of protons [11].

Figure 11 shows the cyclotron and beam simulations related to the particle orbits in the machine. In particular, clean extraction of the high-energy beam is a great challenge. Good turn separation is mandatory in order to place a thin septum inside of the last turn to provide an electrostatic kick to the beam enabling it to exit the machine. Beam loss on the septum must be less than the required part-in-$10^4$ to not preclude hands-on maintenance. This is difficult because turn separation decreases at outer radii (as 1/r), and very high currents are accompanied by large space-charge forces that tend to make the beam bunch larger. Mitigating these requires the highest-possible energy-gain per turn, in the case of PSI 4 large RF single-gap cavities contribute 500 kV of energy gain each. In addition, very careful management of beam is required to ensure a minimum of emittance growth due to space-charge blowup, particularly at the lower-energy end of the accelerator chain where space-charge effects are most severe.

The exceptional performance achieved by the PSI complex is a tribute to over 40 years of development effort, and the fact that they routinely achieve 99.98% extraction efficiency is a testament to their excellence.

The question remains as to whether this performance could be extended to the higher energies and higher powers required for DAEδALUS. This is being actively explored by Andreas Adelmann and his beam-dynamics group at PSI.

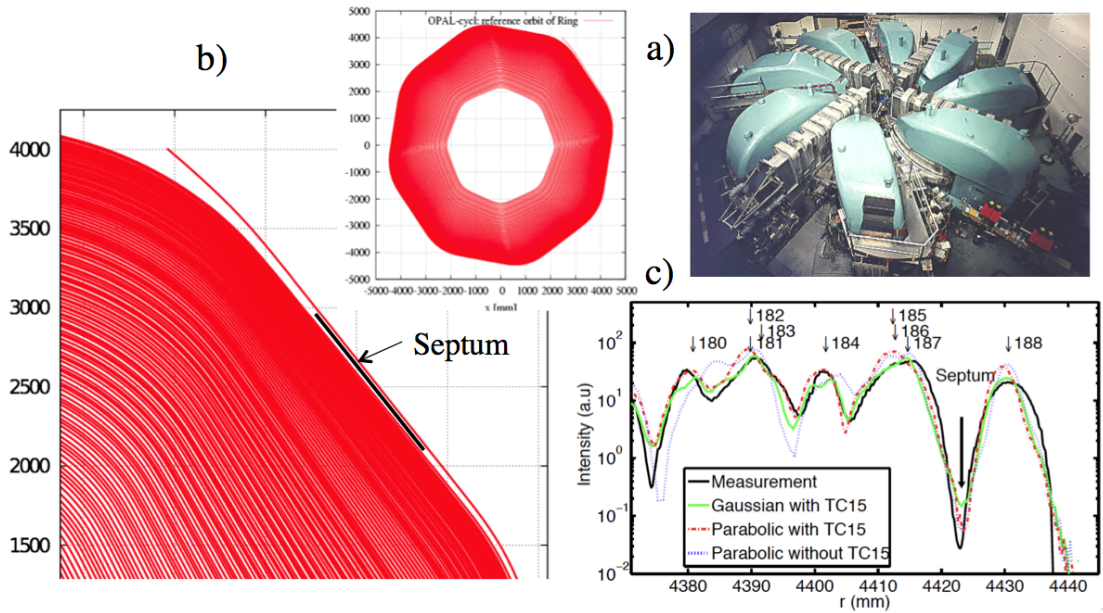

Figure 11: PSI Ring Cyclotron. a) Photograph of cyclotron: 8 (green) magnets are normal-conducting and span about 15 meters. 4 large RF tanks (banded structures) provide 2 MeV/turn energy gain. b) OPAL simulation [13] of turns in the cyclotron, turns get closer together at outer radius. c) A controlled resonance is induced to produce the turn separation needed to cleanly extract the proton beam. The septum intercepts less than 0.02% of the beam, so that 99.98% extraction efficiency is obtained.

## 5.2. $H_2^+$ Cyclotron

A novel concept towards achieving high-current, high-energy beams was developed by L. Calabretta [1] in response to a suggestion by Carlo Rubbia in the 1990's to use high-current cyclotrons for driving thorium reactors [14]. This concept involved the acceleration of $H_2^+$ ions instead of protons, and using a stripping foil to reduce the ion to two protons that could be cleanly extracted from the cyclotron. The higher rigidity of the $H_2^+$ ion ($q/A = 0.5$) compared to bare protons ($q/A = 1$) would require a larger cyclotron, however with the higher fields available from superconducting magnets the size of the machine would still be within reason. In fact, the RIKEN SRC [12] is close to the field and size specifications required.

Stripping extraction has been used extremely effectively for lower-energy cyclotrons that accelerate H- beams, however the lower binding energy (0.7 eV) of the H- ion renders it susceptible to Lorentz stripping in fields as low as 2T and energies below about 70 MeV. The higher binding energy of the $H_2^+$ ion (2.7 eV, at least in its ground state) renders it more stable, and able to survive to 800 MeV in the highest 6T fields anticipated in the DAEδALUS SRC.

The value of stripping extraction is that turn separation is no longer an issue. All ions will pass through the stripper foil, even if turn separation is not clean. Ions will be stripped from several turns, probably not more than 2 or 3, and the protons will carry the energy associated with their turn number. The extraction channel, which will pass through the central region of the SRC (as the protons are bent in rather than out as is the case with H-), will need adequate momentum acceptance to transmit all the protons from whatever turn they are stripped at.

Figure 12 shows the schematic layout of the $H_2^+$ accelerating module, which includes an injector cyclotron that captures up to 5 mA (electrical) of $H_2^+$ and accelerates the beam to about 50 MeV/A. This beam is extracted conventionally for injection into the SRC. This second machine consists of 8 wedge-shaped superconducting magnets and 6 RF cavities (4 of the PSI single-gap type, 2 double-gap for an extra energy boost). The stripper foil is located at the outer radius around 1 o'clock in the figure, at the trailing edge of one of the sector magnets, and the extraction channel comes out roughly along one of the valleys about 270° away.

Figure 12 also shows schematically one of the beam dumps, a graphite block with a hole shaped to correspond to the beam profile so energy is uniformly distributed over a wide area. The graphite is surrounded by a copper, water-cooled jacket and is expected to easily dissipate 6 MW of beam power (see Figure 13).

It is most likely that the power requirements for the DAEδALUS 20 km station will require more than one of these cyclotrons, however adding a second of these relatively compact modules should hopefully prove economical and is certainly feasible.

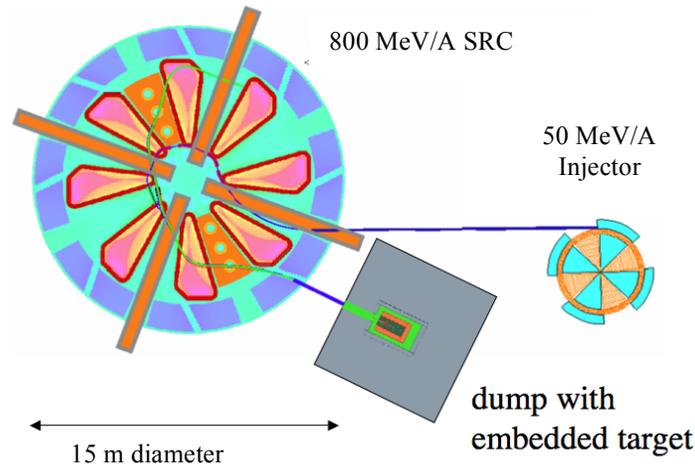

Figure 12: Schematic of $H_2^+$ accelerating module, with an injector cyclotron bringing up to 5 emA of $H_2^+$ ions to about 50 MeV/A. Beam is injected into the SRC and accelerated in about 380 turns to reach 800 MeV/A. A thin carbon stripping foil (~ 1 mg/cm$^2$) reduces the $H_2^+$ to two protons that follow the extraction trajectory through the central region of the SRC and exit into the target.

## 6. Status of Design

The current status of the design effort is summarized in an extensive arXiv report [1], and in two recent IPAC papers [15] [16]. Good progress is being made on many fronts, working towards the goal of confirming feasibility of the overall concept (with a workable conceptual design) and obtaining a first-cut cost estimate. It is expected both of these should be completed within about one year. A brief synopsis of each major component will be given here.

### 6.1. Ion Source

Most any ion source for protons will also produce $H_2^+$ ions, with yields that can be tuned to be as high as 85% [17]. Currents of the order of the 30 mA required are within reason, at emittances suitable for cyclotron injection. The VIS microwave source at the LNS-Catania [18] is being made available for tests.

One challenge is to control the production of loosely-bound vibrational states. $H_2^+$ ions are naturally formed in a range of states, 17 are bound, with very long lifetimes [19], pretty much independent of the ion source plasma characteristics. Population follows a Franck-Condon distribution, with about 10% in $\nu \geq 8$ states, which have binding energies low enough to be at risk of Lorentz stripping in the high fields of the SRC. Studies have shown that introduction of noble gases (He, Ne) into the source plasma can collisionally quench loosely-bound states [20], this must be tested to determine if a suitable level of extinguishing of these potential sources of beam loss can be obtained. Laser diagnostics are capable of this[2]. In addition a laser-based cleaning technique developed at Oak Ridge [21] has potential for dissociating loosely-bound states, with extremely high efficiency.

### 6.2. Injector Cyclotron

The $H_2^+$ ions are axially injected into the central region of the first cyclotron [16] with a classical spiral inflector. Design of the central region to capture the required 5 mA is challenging, the required capture efficiency is high by normal cyclotron standards, and more so considering the high space-charge effects anticipated. A collaboration with BEST Cyclotrons [22] has been established to use their existing test stand to optimize the design of the central region. The Catania VIS source will be shipped to Vancouver for these tests.

---

[2] Von Busch et al (ref [19]) perform dissociation with a Hg flash-lamp. Lasers available today (modest-power NdYAG or TiSapph) would produce significantly higher dissociation yields.

Beam dynamics for the Injector Cyclotron, and basic engineering details for the magnet, RF and the conventional electrostatic-septum-based extraction system, have been studied by the LNS-Catania group [16], showing no really difficult technical issues.

### 6.3. Superconducting Ring Cyclotron (SRC)

Beam dynamics studies for the SRC are being carried out at LNS-Catania [15] and PSI [13]. Achieving isochronicity, vertical focusing and safe trajectories in tune-space (avoiding beam blowup via resonance crossing), while maintaining a feasible design for the magnet steel and superconducting coils is a challenging endeavor, but is converging steadily. Details can be found in [1] and [15]. Good design concepts for the RF cavities have been developed, as have feasible solutions for the injection and extraction orbits. Vacuum specifications for minimizing gas-stripping beam losses have been assessed, and while challenging ($\sim 10^{-9}$ torr), are within engineering feasibility. Overall, it does appear that the required beam-loss specification can be met.

A search is being conducted to select a superconducting magnet designer to work with the beam-dynamics team, to develop an optimum design that satisfies the beam-physics requirements and yields a magnet design that is feasible, buildable, and cost-optimal. Currently the parameters for the magnet, in terms of maximum field ($\sim 6.3$T), hoop forces (>500 kN), as well as the cryostat design, are presenting "interesting" engineering challenges, but are felt to not be without acceptable solutions.

A workshop is being held in Erice in early December 2011 that will gather cyclotron experts to study the complex issues with this cyclotron, and is expected to chart a course for developing a feasible design and roadmap for future progress.

### 6.4. Target

A target design capable of absorbing 6 MW of 800 MeV protons has been developed [23], using a re-entrant design shown in Figure 13. Thermal shock effects from pulsed beam are being studied, and may influence the recommended beam-on/beam-off cycles for the accelerators.

Code studies are being conducted using MARS [24] to evaluate pion production and confirm that the graphite target and 800 MeV protons represents an optimum for the experiment.

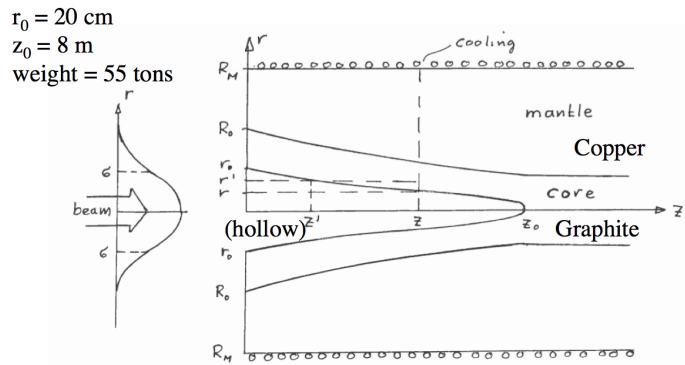

Figure 13: Concept for 6 MW target with representative materials and dimensions. Sketch is not to scale.

### 7. Outlook

Decay-at-rest neutrino sources offer exceptional opportunities for addressing topics of great interest in neutrino physics today. Flexibility in placement of compact sources at different baseline lengths from a suitable detector is a new degree of freedom previously unavailable for this research. Key is obtaining compact, cost-effective sources suitable for these experiments.

The concept for $H_2^+$ acceleration, with potential for producing very high power beams while keeping beam losses to acceptably low levels, is viewed as a viable path towards this capability. The current design effort is proceeding at a very satisfactory pace, so far no serious problems have been uncovered that would render the concept not possible or practical. It is expected that within about a year the feasibility of this approach should be established, and a first cost be available for an accelerator module.

It should be added that this accelerator concept could well represent a breakthrough in price-performance parameters for high-power proton beams, and as such become truly enabling technology for thorium reactors, waste burning, or other ADS applications.

## Ackowledgments

Credit for many of the graphics used is owed to Janet Conrad, Michael Shaevitz, Luciano Calabretta, Boris Kayser, Enectali Figueroa-Feliciano, Chris Tschalaer and Georgia Karagiorgi. Author is grateful for these great illustrations, as well as fruitful discussions with these colleagues.